\documentclass[twocolumn,aps,prm,reprint,superscriptaddress,showpacs,floatfix,amsmath]{revtex4-1}

\usepackage{epsfig}
\usepackage{amsmath}
\usepackage{bm}
\usepackage{float}
\usepackage{graphicx}
\usepackage{xcolor}

\usepackage[normalem]{ulem}
\usepackage[T1]{fontenc} 

\begin{document}


\title{Anomalies in the electronic structure of a 5$d$ transition metal oxide, IrO$_2$}

\author{Swapnil Patil}
\affiliation{Department of Condensed Matter Physics and Materials Science, Tata Institute of Fundamental Research, Homi Bhabha Road, Colaba, Mumbai-400005, India}
\affiliation{Department of Physics, Indian Institute of Technology (Banaras Hindu University), Varanasi-221005, India}
\author{Aniket~Maiti}
\affiliation{Department of Condensed Matter Physics and Materials Science, Tata Institute of Fundamental Research, Homi Bhabha Road, Colaba, Mumbai-400005, India}
\author{Surajit~Dutta}
\affiliation{Department of Condensed Matter Physics and Materials Science, Tata Institute of Fundamental Research, Homi Bhabha Road, Colaba, Mumbai-400005, India}
\author{Khadiza~Ali}
\affiliation{Department of Condensed Matter Physics and Materials Science, Tata Institute of Fundamental Research, Homi Bhabha Road, Colaba, Mumbai-400005, India}
\author{Pramita Mishra}
\affiliation{Department of Condensed Matter Physics and Materials Science, Tata Institute of Fundamental Research, Homi Bhabha Road, Colaba, Mumbai-400005, India}
\author{Ram Prakash Pandeya}
\affiliation{Department of Condensed Matter Physics and Materials Science, Tata Institute of Fundamental Research, Homi Bhabha Road, Colaba, Mumbai-400005, India}
\author{Arindam Pramanik}
\affiliation{Department of Condensed Matter Physics and Materials Science, Tata Institute of Fundamental Research, Homi Bhabha Road, Colaba, Mumbai-400005, India}
\author{Sawani Datta}
\affiliation{Department of Condensed Matter Physics and Materials Science, Tata Institute of Fundamental Research, Homi Bhabha Road, Colaba, Mumbai-400005, India}
\author{Srinivas~C.~Kandukuri}
\affiliation{Department of Condensed Matter Physics and Materials Science, Tata Institute of Fundamental Research, Homi Bhabha Road, Colaba, Mumbai-400005, India}
\author{Kalobaran~Maiti}
\altaffiliation{Corresponding author: kbmaiti@tifr.res.in}
\affiliation{Department of Condensed Matter Physics and Materials Science, Tata Institute of Fundamental Research, Homi Bhabha Road, Colaba, Mumbai-400005, India}

\date{\today}


\begin{abstract}
Ir-based materials have drawn much attention due to the observation of insulating phase believed to be driven by spin-orbit coupling while Ir 5$d$ states are expected to be weakly correlated due to their large orbital extensions. IrO$_2$, a simple binary material, shows metallic ground state which seems to deviate from the behavior of most other Ir-based materials and varied predictions in these material class. We studied the electronic structure of IrO$_2$ at different temperatures employing high resolution photoemission spectroscopy with photon energies spanning from ultraviolet to hard $x$-ray range. Experimental spectra exhibit a signature of enhancement of Ir-O covalency in the bulk compared to the surface electronic structure. The branching ratio of the spin-orbit split Ir core level peaks is found to be larger than its atomic values and it enhances further in the bulk electronic structure. Such deviation from the atomic description of the core level spectroscopy manifests the enhancement of the orbital moment due to the solid state effects. The valence band spectra could be captured well within the density functional theory. The photon energy dependence of the features in the valence band spectra and their comparison with the calculated results show dominant Ir 5$d$ character of the features near the Fermi level; O 2$p$ peaks appear at higher binding energies. Interestingly, the O 2$p$ contributions of the feature at the Fermi level is significant and it enhances at low temperatures. This reveals an orbital selective enhancement of the covalency with cooling which is an evidence against purely spin-orbit coupling based scenario proposed for these systems.
\end{abstract}




\maketitle


\section{Introduction}

Transition metal oxides have been drawing much attention during the past few decades due to their exotic properties arising primarily from the interplay of electron correlation ($U$), covalency, spin-orbit coupling (SOC) and coupling with various collective degrees of freedom \cite{Georges,Imada}. With the increase in radial extension of the $d$ orbitals, the correlation strength among them is expected to reduce and subsequently the other degrees of freedom become more dominant. In that respect, 4$d$ oxides also show plethora of interesting properties \cite{4dTMO}. Materials in the 5$d$ family provide additional interests due to the strong SOC. For example,
Ir based compounds are fascinating due to a variety of physical properties exhibited by them like
Mott insulators \cite{Gretarsson}, signature of density wave in insulating phase \cite{BaIrO3}, topological insulators \cite{Yang,Kargarian,Y2Ir2O7}, Weyl semimetals and axion insulators \cite{Wan}, etc. Such varied behavior is ascribed to the peculiar nature of the Ir 5$d$ electrons which are believed to lie at the border between highly correlated Mott state of 3$d$ transition metal electrons and the itinerant behavior exhibited by the mobile $sp$ conduction electrons. Various competing interactions such as crystalline electric field (CEF) splitting, SOC, $U$ compete with each other giving rise to diverse exotic properties. To understand such behavior of the Ir 5$d$ electrons, it is important to study them under reduced crystalline complexity. Since the aforementioned interactions are dominated by the on-site interactions (local), it is expected that such a study will manifest the intended behavior of Ir 5$d$ electrons. IrO$_2$ is one such simple binary compound and offers a good platform for such an investigations. Ir$^{4+}$ has 5$d^5$ electronic configuration. While IrO$_2$ has been drawing attention for long \cite{Werthein-PRB80,Mattheiss_PRB76,Riga}, there is a recent revival of interest in the study of the properties of this compound emphasizing importance of spin-orbit coupling in this material \cite{Miao,Clancy,Hirata}.

\begin{figure}
\vspace{-6ex}
\includegraphics[width=0.9\linewidth]{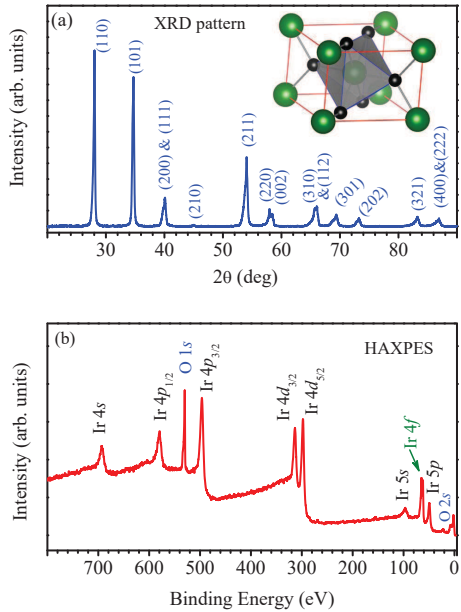}
\vspace{-4ex}
\caption{(a) Powder $x$-ray diffraction (XRD) pattern of IrO$_2$ exhibiting single phase without any impurity peak. The crystal lattice of IrO$_2$ (rutile structure) is shown in the inset. IrO$_6$ octahedron is shown at the center of the unit cell. (b) Survey scan in the wide energy range collected at room temperature using hard $x$-ray photon energy.}
\label{Fig1Str}
\end{figure}

IrO$_2$ forms in a rutile structure with space group $P4_2/mnm$ and lattice constants $a$ = 4.5049 \AA\ and $c$ = 3.1587 \AA. Ir-atoms form a body centered tetragonal unit cell and each Ir atom is coordinated by 6 oxygen atoms forming a distorted octahedron as shown in the inset of Fig. \ref{Fig1Str}(a). IrO$_6$ units have 4 Ir-O bondlengths of 2 \AA\ and 2 somewhat shorter bonds of 1.96 \AA. Under the influence of CEF, the Ir ground state degeneracy is lifted and Ir 5$d$ bands form $t_{2g}$ and $e_g$ bands with $t_{2g}$ contributions lying at a lower energy. The ground state electronic configuration would be $t_{2g}^5$. Earlier works proposed major role of SOC in its electronic structure \cite{Kim,Panda,YPing}. Considering the effective orbital moment, $L_{eff} = 1$ the degenerate $t_{2g}$ bands spit into two bands with effective total angular momentum, $J_{eff}$ = 3/2 and 1/2. Thus, the bands crossing the Fermi level, $\epsilon_F$ largely possess $J_{eff}$ = 1/2 character and derives the electronic properties of this system. In parallel, another study \cite{Kahk-PRL14} suggested that crystal field splitting of the bands due to the distortion of the IrO$_6$ octahedron lifts the degeneracy of the $t_{2g}$ bands and hence, the description based on $L_{eff} = 1$ for the $t_{2g}$ orbitals may not be applicable here. Subsequently, an angle resolved photoemission study suggested importance of spin-orbit coupling in the electronic properties \cite{PranabDas_PRM18}. Clearly, the electronic structure of IrO$_2$ is an outstanding puzzle and it is important to find out the scenario experimentally on the competing spin-orbit coupling and crystal field induced effects in this material.

We report here our results of the investigation of the electronic structure employing high resolution photoemission spectroscopy with a variation of surface sensitivity and temperatures. We analyze our results within the framework of the correlated electron systems \cite{Georges,Imada,Patil_JPCM} and discover signatures of deviation from typical description of the electronic structure for such systems.

\section{Methods}

The IrO$_2$ sample was prepared from commercially available high quality (purity of 99.9\%) IrO$_2$ powders, which were pelletized using a pressure of 5 tonnes and then sintered at 800 $^o$C for 3 days in air to get well sintered hard pellet with large grain size. The structural and elemental characterization of the sample was done by $x$-ray power diffraction (XRD) and energy dispersive analysis of $x$-rays (EDX). A typical XRD pattern is shown in Fig. \ref{Fig1Str}(a) exhibiting single phase. We did not find traces of foreign element in our sample.

The photoemission spectroscopy was carried out using a state-of-the-art high resolution spectrometer equipped with the VG-Scienta R4000 hemispherical electron energy analyzer and monochromatic photon sources. The energy resolution for He {\scriptsize I} ($h\nu$ = 21.2 eV) and He {\scriptsize II} ($h\nu$ = 40.8 eV) spectra was fixed at 5 meV. For conventional $x$-ray photoemission spectroscopy (CXPS), we used a monochromatic Al $K\alpha$ ($h\nu$ = 1486.6 eV) source; total energy resolution for these measurements was set to 300 meV. In order to enhance the bulk sensitivity of the technique, we carried out hard $x$-ray photoemission spectroscopy (HXPS)\cite{HAXPES} at the synchrotron facility, PETRA III DESY, Hamburg, Germany. The photon energy used was ($h\nu$ =) 5947.6~eV and the energy resolution was fixed at 150 meV. The sample surfaces for the photoemission measurements were prepared by fracturing the sample via top-post removal method to expose the clean surface. The base pressure of the photoemission spectrometer consisting of laboratory sources was 5$\times$10$^{-11}$ torr. The pressure in the HXPS spectrometer was 2$\times$10$^{-10}$ torr during the measurements. The Fermi level, $\epsilon_F$ was derived from the photoemission spectra of high quality Ag mounted in electrical contact with the sample. The temperature variation of the sample was achieved by using an open cycle He cryostat.
Since the sample surfaces probed using different setup can be different and the hard $x$-ray beamline did not have the photon energy below 2.5 keV, we have used the same sample in both sets of experiments, prepared the surface in the ultrahigh vacuum condition and measurements were done immediately after the surface preparation. The reproducibility of the features was verified after each trial of surface preparations. This ensured that the high-resolution data obtained using laboratory sources and synchrotron sources represent the spectral functions of the same sample.

To characterize the features in the experimental spectra, the electronic band structure calculations were carried out using full potential linearized augmented plane wave method within the local spin density approximations as implemented in WEIN2k software \cite{Wien}. The lattice parameters reported by Panda et al.\cite{Panda} have been used in the calculations. The Perdew-Burke-Ernzerhof functional revised for solids (PBEsol)\cite{perdew} exchange-correlation functional was used and the formulation of Anisimov {\it et al.}\cite{anisimov} was adopted to include the effective Coulomb interaction, $U_{eff}$ among Ir-5$d$ electrons. We found that the calculated results for various $U_{eff}$ values between 0 - 2 eV provide quite similar descriptions apart from a small energy shift of the features consistent with earlier observations \cite{Panda,YPing,PranabDas_PRM18}. The results shown here correspond to the calculations including SOC and $U_{eff}$ = 2 eV. The convergence was achieved by considering 1000 $k$ points within the first Brillouin zone.

\section{Results and Discussions}

To investigate the quality of the sample, a survey scan was collected using HXPS and CXPS. Since HXPS has significantly large bulk sensitivity (escape depth, $\lambda$ of valence electrons is about 40 \AA), it will reveal the bulk electronic structure of the sample. The experimental HXPS results are shown in Fig. \ref{Fig1Str}(b). All the peaks correspond to various core levels of Ir and O establishing high purity of the sample.

\begin{figure}
\vspace{-2ex}
\centering
\includegraphics[width=0.9\linewidth]{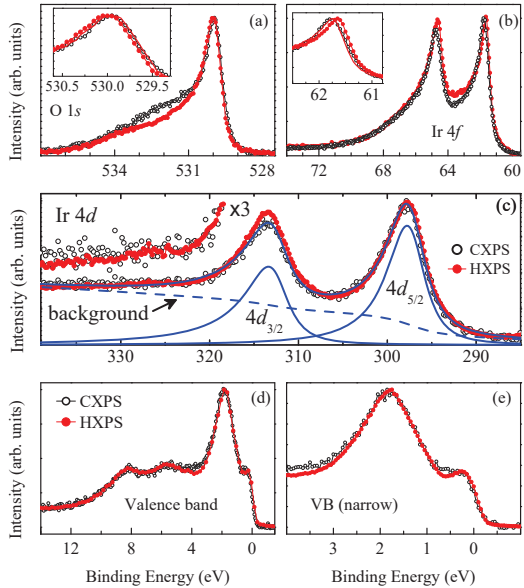}
\vspace{-8ex}
\caption{(a) O 1$s$, (b) Ir 4$f$, (c) Ir 4$d$, (d) valence band (wide) and (e) valence band (close to the Fermi level) spectra collected using conventional XPS (CXPS: open circles) and HXPS (solid circles). Insets in (a) and (b) show the most intense peak in an expanded energy scale. The lines in the insets are the shifted HXPS data providing an energy shift of 100 meV of Ir 4$f$ peak and 50 meV shift of O 1$s$ peak. The lines in (c) are the fit results of Ir 4$d$ CXPS data. The higher binding energy region of Ir 4$d$ spectra is shown in an enhanced intensity scale.}
\label{Fig2-core}
\end{figure}

The surface sensitivity of the photoemission spectroscopy technique becomes most prominent with the smallest escape depth of photoelectrons of about 6 \AA\ for the electron kinetic energy 40 - 100 eV \cite{universalCurve}. Since, the electronic structure at the surface can be very different from the bulk electronic structure \cite{surf_1}, it is important to verify if such an effect is also present in IrO$_2$. In Fig. \ref{Fig2-core}, we plot the core level and valence band spectra collected using conventional $x$-ray source of Al $K\alpha$ and hard $x$-rays from synchrotron. The oxygen 1$s$ spectra shown in Fig. \ref{Fig2-core}(a) exhibit a sharp peak at about 530 eV binding energy along with a feature at 532.4 eV leading to an asymmetry in the lineshape. The intensity at 532.4 eV reduces significantly in the HXPS spectrum due to the increase in bulk sensitivity ($\lambda\sim$38 \AA\ for HXPS and $\sim$16 \AA\ for CXPS) manifesting their origin linked to the surface electronic structure. These oxygens could be the surface oxygens, less bound compared to the bulk ones and/or the oxygens adsorbed on the surface. We observe strong asymmetry in the lineshape arising from the metallicity of the system having significant density of states at the Fermi level allowing excitations across the Fermi level along with the core level photoemission. Interestingly, the peak position is shifted towards higher binding energies in HXPS data [see inset of Fig. \ref{Fig2-core}(a)] which is a signature of an enhancement of local potential (Madelung potential) at the oxygen sites in the bulk. The energy shift is estimated to be 50 meV via shifting the HXPS data (line in the inset) and superimposing over the CXPS data.

Ir 4$f$ spectra are shown in Fig. \ref{Fig2-core}(b) exhibiting two sharp peaks at 61.8 eV and 64.8 eV binding energies in the CXPS data ($\lambda\sim$19.6 \AA) indicating a spin-orbit splitting of about 3 eV for the 4$f$ states. In the bulk sensitive HXPS spectra ($\lambda\sim$39.9 \AA), the peak position is shifted by 100 meV towards lower binding energy; energy shift is estimated by shifting the HXPS data as shown by a line in the inset of Fig. \ref{Fig2-core}(b). Such an energy shift towards lower binding energy for the cation and the shift of the O 1$s$ peak towards higher binding energy is a signature of an enhancement of covalency (decrease in ionicity) in the bulk electronic structure \cite{Kerber-JVac96}. Such a scenario may be expected as the bulk contains complete periodicity of the solid with complete IrO$_6$ octahedra while surface often has uncompensated bonds.

In addition, we observe an increase of the 4$f_{5/2}$ peak intensity relative to the intensity of 4$f_{7/2}$ peak in the HXPS data. Thus, the branching ratio (the ratio of the intensities of Ir 4$f_{5/2}$ and Ir 4$f_{7/2}$ peaks) is enhanced in the HXPS data which can happen due to an enhancement of the effective orbital moment in the bulk. Ir 4$d$ spectra shown in Fig. \ref{Fig2-core}(c) exhibit two intense features at 298 eV and 313.5 eV binding energies corresponding to 4$d_{5/2}$ and 4$d_{3/2}$ photoemission, respectively. We do not observe discernible features associated to the satellite signal often observed in such core level spectroscopy due to electron-electron Coulomb repulsion. In order to observe this better, the higher energy part is shown in an enhanced intensity scale exhibiting signature of very weak features. All these results demonstrate that the correlation effect, if there is any, will be weak in this system as expected for 5$d$ orbitals constituting the valence bands having large orbital extensions. Interestingly, Ir 3$d$ signal also shows enhancement of the 3$d_{3/2}$ peak intensity relative to the 3$d_{5/2}$ peak in the HXPS spectrum ($\lambda\sim$39 \AA) compared to the CXPS data ($\lambda\sim$17.9 \AA).

The total angular momentum, $J_c$ of the core hole can be expressed as $J_c = L_c \pm S_c$ where $L_c$ and $S_c$ are the orbital and spin quantum numbers of the core hole, respectively. Considering the intensity of the peaks scales with the multiplicity of the eigenstates, the branching ratio for Ir 4$f$ will be $[2(L_c-S_c)+1]:[2(L_c+S_c)+1]$ which is 3:4 and the branching ratio for Ir 4$d$ would be 2:3. We have estimated the branching ratio by fitting the core level spectra using standard peak fitting procedure following the least square error method.
We have used Shirley background function and the peak shapes were represented by \emph{Pearson IV} functions to capture the asymmetry, experimental resolution broadening and lifetime broadening.
A typical fit is shown by lines in Fig. \ref{Fig2-core}(c) for the case of Ir 4$d$ CXPS data. The estimated branching ratio for 4$d$ signal is found to be 0.81$\pm$0.05 for CXPS data and 0.85$\pm$0.05 for HXPS data which is significantly higher than the atomic value of 0.67. For the 4$f$ signal, this ratio is found to be 0.86$\pm$0.05 for the CXPS data and 0.88$\pm$0.05 for the HXPS data in contrast to an expectation of 0.75. While the atomic description often found to be good in various core level spectroscopy \cite{dimen}, a deviation from such a behavior may occur due to the solid state effects \cite{Arindam}. The spin-orbit coupling strength in an external scalar potential, $V(r)$ can be expressed as $-{{e\hbar}\over{(2mc)^2}}\sigma.(E(r)\times p)$. Here, $E(r)$ is the electric field (= -$\nabla V(r)$) and $p$ is electron momentum. In IrO$_2$, $E(r)$ becomes significant presumably due to the uniaxial anisotropy in the rutile structure. Thus, the observation of an enhancement of branching ratio in the bulk electronic structure is attributed to various solid state effects including asymmetries of the IrO$_6$ octahedra.

In Fig. \ref{Fig2-core}(d), we show the CXPS and HXPS data for the valence band. Both the spectra looks almost exactly identical; a small difference in linewidth may occur due to the energy resolution. This shows that the valence band in the surface and bulk electronic structure are not very different and/or the change in the spectral function is below the sensitivity of the technique. The near Fermi level region is shown in Fig. \ref{Fig2-core}(e) in an expanded scale. Clearly, the HXPS data show distinctly defined peaks due to the better energy resolution relative to the laboratory $x$-ray source experiments.

\begin{figure}
\vspace{-2ex}
\centering
\includegraphics[width=0.9\linewidth]{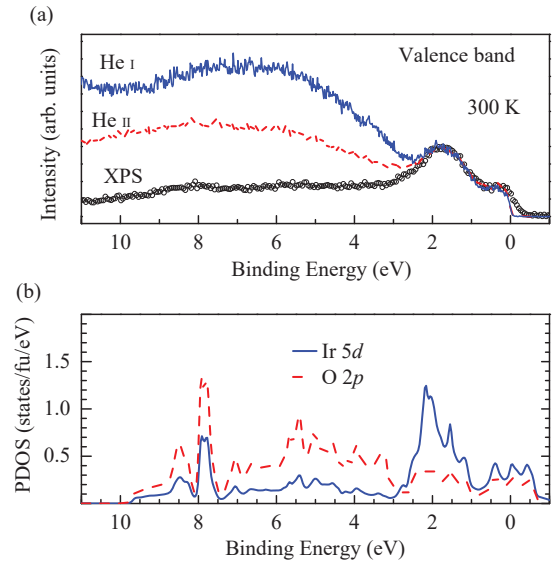}
\vspace{-8ex}
\caption{(a) Valence band spectra at room temperature collected using photon energies, Al $K\alpha$ (open circles), He {\scriptsize II} (dashed line) and He {\scriptsize I} (solid line). (b) The calculated partial density of states of Ir 5$d$ (solid line) and O 2$p$ (dashed line) states.}
\label{Fig3-VBW}
\end{figure}

The valence band spectra are investigated in Fig. \ref{Fig3-VBW}(a), where the data collected using He {\scriptsize I}, He {\scriptsize II} and Al $K\alpha$ photon energies are superimposed over each other after normalization by the intensities near $\epsilon_F$. All the spectra show two distinct features close to the Fermi level with comparable intensities and width. There is a strong monotonic enhancement in intensity beyond about 3 eV binding energy with the decrease in photon energy. The photoemission cross-section \cite{yeh} of Ir 5$d$ states are 24.29, 23.66 and 0.016, and the cross-section of O 2$p$ states are 10.67, 5.816 and 0.00023 for photon energies 21.2 eV, 40.8 eV and 1486.6 eV, respectively. This makes the ratio of cross-sections (O 2$p$/Ir 5$d$) to be 0.439, 0.246 and 0.014. Clearly, the relative intensity of O 2$p$ signal is the highest in He {\scriptsize I} spectrum and becomes insignificant at Al $K\alpha$ photon energy. This suggests that the electronic states constituting the two features A and B possess Ir 5$d$ orbital character. The broad features beyond 3 eV largely comprises of O~2$p$ states hybridized with the Ir~5$d$ states whose photoemission intensity gets enhanced at He~{\scriptsize I} due to cross-section effects \cite{yeh} consistent with the description in other transition metal based systems \cite{ruthenates}.

In order to verify these assertions, we compare the experimental features with the calculated density of states in Fig. \ref{Fig3-VBW}(b). The Ir 5$d$ partial density of states (PDOS) shown by solid line exhibit several features with highest intensity in the vicinity of $\epsilon_F$. Clearly the feature at 2 eV has the highest Ir 5$d$ $t_{2g}$ character. The intensities beyond 3 eV is dominated by O 2$p$ PDOS contributions. Near 8 eV binding energy, Ir 5$d$ PDOS seem to peak although the dominant contribution is still from O 2$p$ PDOS. In fact, we observe Ir 5$d$ contributions spanning over the whole energy range due to the presence of $\pi$ and $\sigma$ bonding contributions, respectively. All these results reflect strong covalent nature of the system and are consistent with the experimental observations.

The relative intensities of the two distinct features near $\epsilon_F$ remains roughly the same across the photon energies used; a small change at the Fermi level is seen due to the resolution broadening. It is to note here that the feature near $\epsilon_F$ possess stronger mixed character than the feature at about 2 eV binding energy. Such difference in the characters of the states manifests their independence in properties.

Considering that the features beyond 3 eV represent the contributions from O 2$p$ - Ir 5$d$ bonding of $\pi$ and $\sigma$ type as well as non-bonding O 2$p$ intensities, it is tempting to attribute the features near $\epsilon_F$ to $J_{eff}$ = 1/2 and $J_{eff}$ = 3/2 scenario as often described in the literature \cite{Riga,Panda}. It is however to note here that the calculations without considering spin-orbit coupling provides a good description of these features. Therefore, the necessity to bring in the $J_{eff}$ scenario which requires strong spin-orbit coupling of degenerate $t_{2g}$ bands is debatable although this scenario also provides similar features. Moreover, the distortion of the IrO$_6$ octahedra lifts the degeneracy of the $t_{2g}$ bands \cite{Kahk-PRL14}.

\begin{figure}
\vspace{-2ex}
\centering
\includegraphics[width=0.9\linewidth]{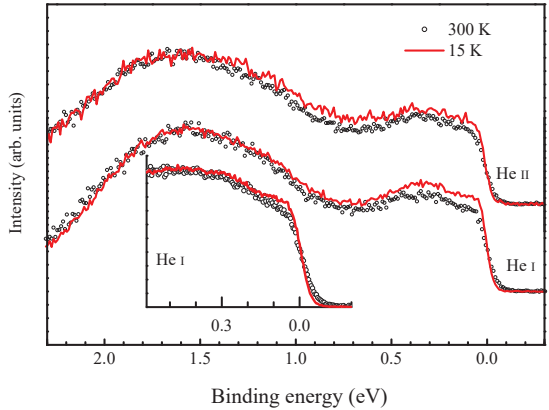}
\vspace{-28ex}
\caption{He~{\scriptsize I} and He~{\scriptsize II} valence band spectra at 300 K (open circles) and 15 K (solid line). The inset shows the near Fermi level part of the He {\scriptsize I} spectra after normalizing by the intensities at 0.5 eV.}
\label{Fig4-VBN}
\end{figure}

So far, we discussed the identification of various features and their association to surface-bulk electronic structures. We now investigate the temperature dependence of the features close to $\epsilon_F$ collected with high resolution. In Fig. \ref{Fig4-VBN}, we show the He {\scriptsize I} and He {\scriptsize II} spectra collected at 15 K (solid line) and 300 K (open circles). There are two broad spectral regions; one close to $\epsilon_F$ and the other beyond 0.7 eV binding energy. The features at higher energies are essentially Ir 5$d$ states having $d_{xz}$, $d_{yz}$ and $d_{x^2-y^2}$ symmetry as reported earlier \cite{Kahk-PRL14}. The intensities near $\epsilon_F$ possess $d_{xz}$ and $d_{yz}$ symmetry along with a significant contributions from O 2$p$ states \cite{yeh}. A normalization of the intensities by the intensity around 1.5 eV peak exhibit significant enhancement of intensities in the vicinity of $\epsilon_F$ at low temperature. While there appears to be an overall increase in intensity in the He {\scriptsize II} spectra, distinct changes are observed in the He {\scriptsize I} spectra exhibiting large enhancement of the near $\epsilon_F$ features. Thus, there are two possible conclusions - (i) this enhancement is linked to O 2$p$ contributions and (ii) the states having $d_{xz}$ and $d_{yz}$ have different behavior than the $d_{x^2-y^2}$ states. It is to note here that the apical oxygens (on $z$-axis) are closer to the Ir sites (smaller bondlength) compared to those in the $xy$ plane. Therefore, $d_{xz}$ / $d_{yz}$ bands will be more covalent due to stronger hybridization. From the experimental results, it is clear that the thermal compression enhances the hybridization differently for $d_{xz}$ / $d_{yz}$ and $d_{x^2-y^2}$ orbitals. Such an orbital selective temperature evolution of the spectral functions provides an evidence against the scenario captured considering an $L_{eff}$ = 1 for degenerate $t_{2g}$ bands.

In the inset, we show the region very close to $\epsilon_F$ after normalizing by the intensity at 0.5 eV. There is a significant enhancement of the intensity near $\epsilon_F$. In a correlated system, the electron energies are strongly influenced by the electron-electron Coulomb repulsion and the corresponding spectral contributions appear away from the Fermi level - these are called Hubbard bands or incoherent features. In photoemission, one probes the lower Hubbard band while the upped Hubbard band appears in the unoccupied part of the electronic structure. There are contributions from the electrons whose energies are not significantly influenced by the electron correlation effects and appear at the Fermi level - this is called a coherent feature which can be captured well using density functional theory. With the reduction in temperature, it appears that the intensity of the coherent feature increases at the cost of the incoherent features as expected.

\begin{figure}
\vspace{-2ex}
\centering
\includegraphics[width=0.9\linewidth]{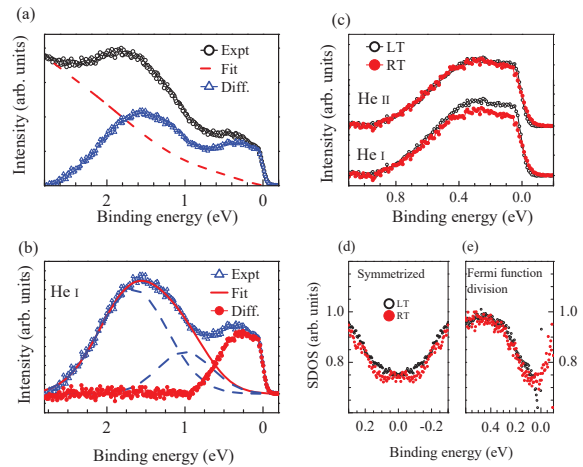}
\vspace{-24ex}
\caption{(a) An example of the subtraction of the tail of higher binding energy contributions. Experimental data (open circles), tail (dashed line) and subtracted data (open triangles). (b) An example of the extraction of the near Fermi level peaks via subtraction of the higher energy features (solid line). Dashed lines are the peaks used to derive higher binding energy contributions. (c) Near Fermi level part from He {\scriptsize I} and He {\scriptsize II} spectra collected at 300 K (solid circles) and 15 K (open circles). (d) Spectral density of states (SDOS) calculated via symmetrization of the He {\scriptsize I} spectra. (e) SDOS extracted by dividing the He {\scriptsize I} spectra by resolution broadened Fermi-Dirac function.}
\label{Fig5-VBNN}
\end{figure}

In order to investigate this further, we extracted the Ir 5$d$ contributions via subtracting tail of the higher binding energy features as shown in Fig. \ref{Fig5-VBNN}(a) for the He {\scriptsize I} room temperature data. The contributions beyond 1 eV is subtracted as shown in Fig. \ref{Fig5-VBNN}(b) - here, the dashed line are the peaks used to simulate the intensities beyond 1 eV binding energy. The spectral contributions extracted in this way are shown in Fig. \ref{Fig5-VBNN}(c). We observe an overall enhancement of the intensities in the whole spectral region shown in Fig. \ref{Fig5-VBNN} in the He {\scriptsize I} spectra; He {\scriptsize II} spectra show marginal change. The larger enhancement in He {\scriptsize I} spectra indicate the change in intensities is linked to the enhancement of oxygen 2$p$ contributions in this spectral region.

The Fermi-Dirac distribution function, expressed as, $F(\epsilon,T) = 1/[exp({{\epsilon-\epsilon_F}\over{k_BT}}) + 1]$ changes the spectral intensities in the vicinity of $\epsilon_F$ with temperature. Therefore, a direct comparison of the raw data at different temperatures near $\epsilon_F$ is difficult. The symmetrization of the Fermi-Dirac function with respect to $\epsilon_F$ (= 0 in the binding energy scale) gives unity; $F(\epsilon,T) + F(-\epsilon,T) = 1$. The experimental spectral intensity can be expressed as $I(\epsilon,T) = SDOS(\epsilon,T)\times F(\epsilon,T)$, where `SDOS' represents the spectral density of states. Thus, one can extract SDOS by symmetrizing the experimental data as follows, $SDOS(\epsilon,T) = I(\epsilon,T) + I(-\epsilon,T)$ assuming that $SDOS(\epsilon,T)$ is symmetric with respect to $\epsilon_F$. Independent of the properties of SDOS, the extracted intensity at $\epsilon_F$ is robust and does not have influence from temperature dependent Fermi-Dirac function. Thus extracted SDOS from the He {\scriptsize I} data are shown in Fig. \ref{Fig5-VBNN}(d). The results at 300 K and 15 K exhibit almost identical intensity at the Fermi level with very small enhancement at about 70 meV at 15 K.

It is, however, not a priori clear if the spectral function and/or the density of states of IrO$_2$ is symmetric with respect to $\epsilon_F$. To verify this, we have divided the experimental spectra by the resolution broadened Fermi-Dirac distribution function as follows, $SDOS(\epsilon,T) = I(\epsilon,T)/F(\epsilon,T)$. Since the experimental resolution is good (5 meV), such an estimation provides a good representation of SDOS. The extracted data are shown in Fig. \ref{Fig5-VBNN}(e). Evidently, the SDOS at both 300 K and 15 K look very similar with small enhancement just below the Fermi level as seen in the symmetrized data. Moreover, we observe that SDOS at 300 K looks symmetric with respect to $\epsilon_F$ providing confidence on the symmetrization process adopted in Fig. \ref{Fig5-VBNN}(d).
Since the spectra near $\epsilon_F$ obtained by the Fermi function division method is noisy due to weak intensities, the symmetrized data is used to obtain an estimate of the spectral function close to $\epsilon_F$.
All these results establish that the correlation induced effect, if there is any, is significantly weak in this system. This is probably the reason for the metallic ground state of this system.

\section{Conclusions}

In summary, we studied the electronic structure of IrO$_2$ employing high resolution photoemission spectroscopy. A combination of hard $x$-ray and conventional $x$-ray data helped to reveal the surface-bulk differences in the electronic structure. The branching ratio of the Ir core level spin-orbit split peaks is significantly larger than their atomic values and it enhances in the bulk. This is attributed to an enhancement of the orbital moment due to the solid state effects. Core level spectra exhibit signature of enhanced covalency in the bulk compared to the surface. The valence band spectra exhibit dominant contribution of Ir 5$d$ states close to the Fermi level which is responsible for the electronic properties of this material. The change in surface sensitivity of the technique does not have significant effect in the valence band and the experimental results could be described well by the effective single particle description captured within the density functional calculations. Evidently, electron correlation induced effect is not strong in this system as has also been manifested in the core level spectroscopy. High resolution data exhibit multiple features and interesting temperature evolution. The evolution of the spectral features with temperature and photoemission cross-section show orbital selective changes in the valence band which is an evidence against purely $J_{eff}$ based scenario. The covalency of the states close to the Fermi level enhances with the decrease in temperature and the O 2$p$ states play an important role in deriving the electronic properties of this material.
These results suggests that while the electron-electron Coulomb repulsion reduces with the enhancement in radial extension in higher $d$-systems along with an enhancement of spin-orbit coupling , the covalency and the crystal field effects play important roles leading to exoticity in these materials.

\section{Acknowledgements}

The authors acknowledge financial support under the project no. 12-R\&D-TFR-5.10-0100. KM acknowledges financial assistance from DAE-BRNS, Government of India under the DAE-SRC-OI Award program (Grant No. 21/08/2015-BRNS/35034).

%

%

\end{document}